\documentclass[nofootinbib,amsmath,amssymb]{revtex4}

\usepackage{pstricks}
\usepackage{graphicx}
\usepackage{dcolumn}
\usepackage{bm}

\begin{document}

\newcommand{\locsection}[1]{\setcounter{equation}{0}\section{#1}}
\renewcommand{\theequation}{\thesection.\arabic{equation}}

\def\F{{\bf F}}
\def\A{{\bf A}}
\def\J{{\bf J}}
\def\af{{\bf \alpha}}
\def\beqn{\begin{eqnarray}}
\def\eeqn{\end{eqnarray}}

\def\dspace{\baselineskip = .30in}
\def\beq{\begin{equation}}
\def\eeq{\end{equation}}
\def\bw{\begin{widetext}}
\def\ew{\end{widetext}}
\def\pl{\partial}
\def\na{\nabla}
\def\al{\alpha}
\def\bt{\beta}
\def\Ga{\Gamma}
\def\ga{\gamma}
\def\de{\delta}
\def\De{\Delta}
\def\da{\dagger}
\def\ka{\kappa}
\def\si{\sigma}
\def\Si{\Sigma}
\def\te{\theta}
\def\La{\Lambda}
\def\lam{\lambda}
\def\Om{\Omega}
\def\om{\omega}
\def\ep{\epsilon}
\def\non{\nonumber}
\def\sq{\sqrt}
\def\sqg{\sqrt{G}}
\def\sp{\supset}
\def\sb{\subset}
\def\l{\left (}
\def\r{\right )}
\def\lq{\left [}
\def\rq{\right ]}
\def\fr{\frac}
\def\la{\label}
\def\hs{\hspace}
\def\vs{\vspace}
\def\inf{\infty}
\def\ran{\rangle}
\def\lan{\langle}
\def\ov{\overline}
\def\tl{\tilde}
\def\tm{\times}
\def\lrar{\leftrightarrow}





\vs{1cm}

\title{Degenerate states in the scalar boson spectrum\\
Is the Higgs Boson a Twin ? }

\author{Berthold Stech}
\email{B.Stech@ThPhys.Uni-Heidelberg.DE}

\affiliation{Institut f\"ur Theoretische Physik, Philosophenweg 16, D-69120 Heidelberg, Germany}


\vs{1cm}


\begin{abstract}
The extension of the standard model to  $SU(3)_L\times SU(3)_R \times SU(3)_C$  is considered. Spontaneous symmetry breaking requires two $( 3^* , 3 , 1)$ Higgs field multiplets with a strong hierarchical structure of their vacuum expectation values. An invariant potential is constructed to provide for these vacuum expectation values. This potential  gives masses to all scalar fields apart from the 15 Goldstone bosons. In case there exists  a one-to-one correspondence between the vacuum expectation values of the two field multiplets, the scalar boson spectrum contains degenerate eigenstates. The lowest eigenstate has a mass near 123 GeV  close to the Higgs-like particle discovered at the LHC. In one class of solutions this lowest state is a nearly degenerate twin state. Each member is a superposition of fields from both multiplets with about equal strength. The twins are non identical twins, namely different combinations of  a conventional Higgs and a Higgs field which is not coupled to fermions, only to gauge bosons. A second class of solutions leads again to degenerate states but in this case the state near $123$~GeV remains a single state even for identical low scale vacuum expectation values in both multiplets.
\end{abstract}

\maketitle

\section{Introduction and Scope.}\label{sec:1}

The author's successful prediction \cite{BSHiggs} of the mass of the Higgs-like boson discovered at the LHC \cite{LHC}  may be fortuitous since it involved a speculation about potentials and their normalization. Also, in the literature, similar mass values in the correct energy range have been obtained using quite different speculations; see the compilation in \cite{Schue}. Nevertheless, our own attempt may be worth to be discussed further. It differs strongly from others and needs little theoretical  apparatus. The model is based on the extension of the Glashow-Weinberg-Salam group to
 $SU(3)_L\times SU(3)_R \times SU(3)_C $, the trinification group \cite{Trini} \cite{TriniPak},  which is a subgroup of $E_6$ \cite{E6}. This group can only be unbroken  at and above the scale where the gauge couplings $g_1$ and $g_2$ unite \cite{BZ}. Thus, this model needs the presence of a very high mass scale $M$ of the order of $10^{13}$~GeV. This scale appears to be the right one for producing light Majorana neutrinos in the observed mass range.
The appearance of such a high scale in the same particle representation of a symmetry group is challenging but not unwelcome. At low energies, the corresponding particle mixings still leave a trace in the limit  $M \to \infty$. 
To break the trinification group down to the standard model group and finally to $U_e(1)$, two scalar matrix fields  $H$ and $\tilde H$  transforming  according to $ (3^*, 3)$ with respect to $SU(3)_L \times  SU(3)_R $  are necessary \cite{BZ} \cite{BS}. Thus, we have to deal with 36 real scalar fields. With respect to the $SU(2)_L $ subgroup they form 6 complex Higgs doublets and 6 complex singlets.
  
Our aim is to construct in a phenomenological way a potential for these scalar fields which provides spontaneous symmetry breaking and gives non-zero masses to all fields except for the 15 Goldstone bosons. Our input consists of the vacuum expectation values of these fields. Like in all present theories these fundamental quantities are not yet understood and thus have to be taken in accordance with known facts and restrictions. In this article we are particular interested in the existence of degenerate mass eigenstates. Some degeneracies are
strictly valid due to to CP invariance. Others approximate ones arise from symmetry breaking parameters small compared to $M$.  A particular interesting approximate degeneracy can occur if the  vacuum expectation values (vevs) of the matrix field $ H $ also appear in the vevs of $\tilde H$. 

For the construction of the potential only vacuum expectation values will be used. It is taken to be of the Coleman-Weinberg type \cite{CW}. No explicit mass parameters, like in the tree potential of the standard model, are used except two tiny ones to account for the non-vanishing determinants of the two matrix fields $H$ and $\tilde H$. They break the discrete symmetry 
$ H \Rightarrow - H$~and $\tilde H \Rightarrow - \tilde H$ and the symmetry under a pure phase transformation.  
          
 The  vacuum expectation value  of the first matrix ($H$) can be chosen diagonal. 
The vev of the field $H^1_1$, $v_1= \lan H^1_1\ran$, is related to the top quark mass according to $m_t = g_t ~ v_1$. $  \lan H\ran^2_2 $ is denoted by $ b $ and connected to the bottom quark mass: $m_b= g_b~b$. It is small compared to $v_1$. The  element  $ \lan H\ran^3_3 $ ,  on the other hand,  is huge and can be identified with $M$. The element $v_1$ is necessarily smaller or at most equal to  $v=174$~GeV, the vacuum expectation value of the Higgs field of the standard model. Thus, the matrix field $H$ has Yukawa couplings to the top and to the $b$ quark and also to a very heavy down type quark $D$. 
 By imbedding the scalar multiplet fields in the $E_6$ model (or its trinification subgroup) \cite{BZ}\cite{BS} the vev's $v_1$, $b$ and $M$ are independent of the generations. The generation matrices, which are vev's of flavon fields, occur as factors in front of $H$ and $\tilde H$. 

The second matrix of scalar fields $\tilde H$ has no Yukawa coupling to fermions (except via a heavy singlet state \cite{BZ} \cite{BS}). The vevs of $\tilde H$ are unknown but have to obey important restrictions. This matrix field  needs to have a sizeable vacuum expectation value of the element 
$  \tilde H^3_2 $  \cite{TriniPak,BZ} which we denote by $ M_2$. This vev breaks the left-right symmetry and thus determines the masses of the right handed vector bosons. The experimental limit on right handed currents provides for a lower limit for $M_2$ of the order of $10$~TeV.  Already the presence of $ M $ and $ M_2$ is sufficient to break the trinification group down to the Glashow-Weinberg-Salam group ($\times~ U(1)$). Like $v_1$ ,  the vev of $ \tilde H^1_1$ denoted by $v_2$ is restricted by the known value  of  $v$. The vevs of the second row of $ \tilde H $, $b_2 = \lan \tilde H ^2_2 \ran$ and $b_3 = \lan \tilde H ^2_3 \ran$ are expected to be small like $b$.  In fact, $v_1$, $v_2$ together with $b$, $ b_2$ , $b_3$ have to obey $ v_1^2 +  v_2^2 + b^2 + b_2^2 +b_3^2 = v^2 $. For small $b's$ this relation  reduces to $ v_1^2 +  v_2^2  = v^2 = (174  ~$GeV$)^2$.
These vevs break the  Glashow-Weinberg-Salam group down to the electromagnetic $U_e(1)$ symmetry. The vev $\lan \tilde  H^3_3 \ran =  M_3$ is unknown and not restricted by known facts. It can be large.

 Defining $H = h+i f$  and $\tilde H = \tilde h + i \tilde f $ where $h, \tilde h, f ,\tilde f $  are matrices of real fields we have in general

 \begin{eqnarray}
 \label{VeV}
\langle h \rangle =
\begin{pmatrix}
v_1 & 0 & 0\\
0    & b & 0\\
0    & 0 & M
\end{pmatrix},
\hspace{1cm}
\langle \tilde h \rangle =
\begin{pmatrix}
v_2 & 0 & 0\\
0 & b_2 & b_3\\
0 &  M_2 &M_3 \\
\end{pmatrix}.
 \end{eqnarray}
Here the phases of the vev's of $ \lan H\ran$ are absorbed by the transformation matrices and the fermion fields.  
$ \lan \tilde H\ran$ cannot be diagonalized anymore. We took all its vevs to be real.

The potential  will have its minimum at specific vacuum field configurations chosen from the general form (\ref{VeV}). All first derivatives of this potential have  to vanish at this point. 
The $ 36 \times 36 $ mass matrix obtained from the second derivatives of the potential produces $36$ mass eigenvalues, $15$  of them belonging to massless Goldstone bosons. The latter will become the longitudinal components of  $W^{\pm}$, $Z^0$ and $12$ heavier vector bosons.   We will discuss the mass values and properties of the lightest of the massive scalars. 

The vev's $v_1$ and $ v_2$ in $H$ and $\tilde H$ are responsible for the low scale spontaneous symmetry breaking. If they have the same value  - for instance as a consequence of the same origin of the vevs of  $ H $ and $ \tilde H  $ - the corresponding two scalar bosons can be nearly degenerate. They are then twins. These twins consists of field components from $H$ and also field components from $\tilde H $ which are not directly coupled to quarks and leptons.  A suggested normalization of the potential gives them a mass of $\simeq$ 123 GeV close to the Higgs boson observed at the LHC \cite{LHC}. However, the near mass degeneracy of the twins will only be present if the influence of the remaining states is not severe. 

\section{ Potential formed from Invariants of $H$ and $\tilde H$ separately.}\label{sec:2}

The potential for the $36$ fields has to be formed from $ SU(3)_L \times SU(3)_R$ invariants. In a first step we consider separately invariants of $H$  and  invariants of $\tilde H$.   Invariants which combine $H$ with $\tilde H$ fields will be added later. As mentioned in the introduction tree type mass terms are not included in the construction.  Our construction is based on the vevs of invariants only. They fix the position of the minimum. Up to order $4$ in the fields one remains with the invariants 
(the determinants of $H$ and $\tilde H$ play no role at this stage) 

\beq
\label{J}
J_1 = (Tr[H^\dagger \cdot H])^2 , ~J_2  = 
Tr[ H^\dagger  \cdot H \cdot H^\dagger \cdot H] ~ ~~\text{and}~ ~~ J_3 = (Tr[\tilde H^\dagger \cdot \tilde H])^2,~ J_4 = 
Tr[\tilde H^\dagger  \cdot \tilde H \cdot\tilde H^\dagger \cdot \tilde H] .
\eeq
The potential  $ V(H, \tilde H)$  is a superposition of the $4$ invariants for which all $36$ first derivatives have to vanish at the proposed minimum.  As in \cite {BSHiggs} the superposition of the invariants must necessarily contain  logarithmic functions of the invariants. In order to have only one logarithmic term,  (\ref{VeV})  will be restricted to contain a single large scale only. Therefore, we take   $M_2 $ and  $M_3  $ proportional to the large scale $M$. 

 For the potential we choose a naive phenomenological ansatz with coupling constants $c_i$ for the quartic interactions.

 \begin{eqnarray}
\label{V1}
V_0 = \frac{1}{t}~(c_1 J_1+ c_2 J_2 + c_3 J_3 + c_4 J_4) \\ \nonumber
\text{with}\\  \nonumber
t= \kappa+ \log [\frac{J_1 J_2 J_3 J_4 }{ \lan J_1 \ran \lan J_2 \ran \lan J_3 \ran \lan J_4 \ran }].
\end{eqnarray}
The constant $\kappa$ introduced here allows to take a convenient value for the denominator in the logarithmic term: the product of vevs of the four invariants. 

To get the stationary point at the wanted position one can perform the shift 
 \begin{eqnarray}
\label{shift}
h_1^1 \rightarrow v_1 + h^{1}_1,~ ~h^2_2 \rightarrow b+ h^{2}_2,~~h_3^3 \rightarrow M + h^{3}_3 ,
~~\tilde h_1^1 \rightarrow v_2 + \tilde h^{1}_1, \\
\nonumber
~~\tilde h_2^2 \rightarrow b_2 + \tilde h_2^2,~~\tilde h_2^3 \rightarrow b_3 + \tilde h_2^3,
~~\tilde h^2_3 \rightarrow M_2 + \tilde h^{2}_3, ~~\tilde h^3_3 \rightarrow M_3 + \tilde h^{3}_3. 
\end{eqnarray}
Now the vanishing of all derivatives of $V_0$  is required for the point $H = \tilde H = 0 $ of the shifted fields.

The condition for the vanishing of all first derivatives at the point $ H = \tilde H = 0$ for the shifted fields has a straightforward  solution. It determines the parameter $\kappa$  and   the
coefficients $c_2$, $c_3$,~$c_4$ in terms of  ${c_1}$  :
 \begin{eqnarray}
\label{kappa}
     \kappa = 4,   \quad \quad
c_2= c_1 \frac {\left(b^2+M^2+v_1^2\right)^2}{b^4+M^4+v_1^4},\\
 \nonumber
c_3=c_1 \frac{ \left(b^2+M^2+v_1^2 \right)^2}{\left(b_2^2+b_3^2+v_2^2+ M_2^2+M_3^2 \right)^2}, \\
 \nonumber
c_4=c_1 \frac{ \left(b^2+M^2+v_1^2\right)^2}{ \left(b_2^2+
b_3^2\right)^2+v_2^4+2 (b_2 M_2 +b_3 M_3 )^2+\left(M_2^2+M_3^2\right)^2}.
 \end{eqnarray}
Obviously, for $M_2^2 + M_3^2 = M^2 $, one has $c_2=c_3=c_4=c_1$   in the large $M$ limit. 

The second derivatives of $\frac{1}{2} V_0$ at the point $H = \tilde H=0 $ of the shifted fields provide now for the $36 \times 36$ mass matrix. Degenerate states can be expected in case there is a relation between the vevs of $\tilde H$ with the vevs of $H$. Such a situation may arise by a high scale  spontaneous symmetry breaking fixing simultaneously the vacuum expectation values of both matrix fields. The closest connection is a one-to-one correspondence of all  vevs of $ \tilde H$ with the vevs of $ H $  combined with the requirement 
$\langle \det \tilde H \rangle  = \pm \langle \det \tilde H \rangle$. 
One gets
\begin{eqnarray}
\label{b}
\langle h \rangle =
\begin{pmatrix}
v_1 & 0 & 0\\
0    & b & 0\\
0    & 0 & M
\end{pmatrix},~~~~
\langle \tilde h \rangle =
\begin{pmatrix}
v_1 & 0 & 0\\
0 & 0& \pm b\\
0 &  M & 0\\
\end{pmatrix}.
 \end{eqnarray}
 i.e. $v_2=v_1$,~ $b_2= 0$,~$ b_3= \pm b$,~$ M_2=M$,~$ M_3=0.$
It follows  
 \beq
\label{1-3}
\langle J_3 \rangle = \langle J_1 \rangle~~~\text{and} ~~~\langle J_4 \rangle = \langle J_2 \rangle 
\eeq
suggesting already that mass degeneration will occur. 
In the present section the sign of $b$ in $\langle  \tilde h \rangle$ does play no role. But for latter purposes we will denote the form with the $+$ sign as case $1$, the form with the $-$ sign as case 2. In the  latter $\langle \tilde h \rangle$ differs from $\langle h \rangle $ only  by an $SU(2)_R~ U$-spin rotation with angle $ \frac{\pi}{2}$.  

This correspondence between the vevs of both matrix fields has besides (\ref{1-3}) the consequence $c_3=c_1$,~$c_4 = c_2$. 
The mass matrix and mass eigenvalues resulting from the potential (\ref{V1})  can now easily be obtained.  The masses  are shown here to second order in $v_1$  for large $M$. 
 \begin{eqnarray}
 \label{eigen}
m_1^2 =  c_1 ( v_1^2 + b^2), ~~~ m_2^2 =  c_1 (v_1^2 +b^2), ~~~m_3^2 = c_1~ (4 M^2 + 5 v_1^2 + 5 b^2) ,~~~
m_4^2= c_1~( 4 M^2 + 5 v_1^2 +5 b^2), \\ 
 \nonumber
\hspace{-3cm}   m_i^2 = 0  ~~~~~~ i = 5 ..............36.~~~~~~~~~~~\quad\quad~~~~~~
\end{eqnarray}
There is a clear degeneracy of the two low and the two high mass states. The equality of the low masses  is entirely due to the identification $v_2 \rightarrow v_1 $. It holds (approximately) also in the more general form (\ref{VeV}) provided $M_2^2+M_3^2 = M^2$  and the $b's$ are small compared to $v_1=v_2$. The $32$ mass zero states can be separated into $17 $ states of mass zero and $15$ mass zero Goldstone states. One zero mass state is due to the so far unbroken general phase transformation of $H$ and $\tilde H$. The remaining  $16 $ zero mass states are due to our provisional neglection of invariants which connect the fields in $H$ with the fields in $\tilde H$. Without them $H$ or $\tilde H$ can be independently transformed by $SU(3)_L\times SU(3)_R$ matrices with no change of the invariants. 

Since $b$ is sufficiently small (the b-quark mass is around $3 - 4 $~GeV), it follows from $v_1^2 + v_2^2 = v^2 $ the numerical value $ v_1 = v_2 = \frac{v}{\sqrt{2}} = 123$~GeV, a value quite close to the mass of the Higgs boson found at LHC.

We identify now these degenerate states, these twins, with the Higgs particle of about $125~$GeV found at the LHC. 
 The twins are non-identical twins. They are combinations of the field $h^1_1$, the usual Higgs field which couples to fermions, and the field $\tilde h^1_1$ which has no Yukawa coupling to fermions and only couples to quarks and leptons via its coupling to gauge bosons \cite{BZ,BS}. The mass value obtained at LHC  fixes the coefficient $c_1$ in (\ref{eigen}) to be very close two $1$:  $m_{Higgs} = v_1 \sqrt{ c_1} \simeq 125$~GeV, i.e. $c_1 \simeq 1.03$.

The value $c_1=1$ may be considered to be preferred in a more general but  speculative context, similar to the speculative prediction made in \cite{BSHiggs}: The potential at the minimum,  to order  $M^4$ and $M^2$, yields
\beq
\label{spec}
  V_{0,min}  = c_1~ ( M^4 +  4 M^2 ((h^3_3)^2 +(\tilde h^3_2)^2 ) .
  \eeq
The input vacuum expectation value $M$ in $H$ and $\tilde H$, appears in the potential simply multiplied by the factor  $c_1$, and the factor $ 4~c_1$ multiplying the mass terms in (\ref{spec})  reflects the $4$ masses $M$ in  the $4$ invariants. These  facts suggest to normalize the potential by choosing $c_1 = 1 $.   It then follows $m_{Higgs} = v_1 = \frac{v}{\sqrt{2}} = 123$~GeV. 
 Radiative corrections and in particular the influence of not yet considered invariants combining $H$ and $\tilde H$ can modify this value and lift the degeneracy.  
   
\section{Combining $H$ and $\tilde H$ fields and the mass spectrum.}\label{sec:3}

                The inclusion of invariants neglected up to now is necessary to get an acceptable mass spectrum:  in order not to be in conflict with experiments,  all of the so far $17 $ massless states should become heavier than the standard model like Higgs. This cannot be achieved without problems and caveats.
                
There are quite a number of different invariants containing the fields of both multiplets $H$ and $\tilde H$ \cite{TriniPak}. Most combinations for which all first derivatives vanish lead to some negative eigenvalues of the mass matrix. Also the ones with positive mass values will in general modify the Higgs twin states, mix them with other states and change their masses or have troublesome discontinuities by a change of parameters. We take for the additional potential  the combination
 \beq
  \label{VS} 
\hspace{-5cm} 
V_S = \frac{v_1^2}{M^2} ( r_1 J_1 + r_2 J_2+ r_3 J_3 + r_4 J_4 +   r_5 J_5) + r_6 J_6 + r_7 J_7 + r_8 J_8 + r_9 J_9 \quad \quad \\
\eeq
with the new invariants
 \begin{eqnarray}
\label{J2}
J_5 =  Tr [H^{\dagger}\cdot \tilde H \cdot \tilde H^{\dagger} \cdot H],~~ 
J_6 = Tr [H^{\dagger}\cdot H \cdot \tilde H^{\dagger} \cdot  \tilde H] ,~~
J_7 =  Tr  [H^{\dagger}\cdot \tilde H \cdot  H^{\dagger} \cdot \tilde H] +
Tr  [\tilde H^{\dagger}\cdot H \cdot \tilde H^{\dagger} \cdot H], ~~~\\  \nonumber
J_8 = b~(\det{H} +\det {H^{\dagger}}),~~J_9 = b~( \det{\tilde H} +\det {\tilde H^{\dagger}}).~~~~~~~~~~~\quad\quad~~~~~~~\quad \quad
\end{eqnarray}
The contributions of  $J_1 ...J_5$ turn out to be extremely small compared to the ones for $V_0$ as long as the other $r$ parameter of the new invariants are kept of order 1. For this reason we put the factor $\frac{v_1^2}{M^2}$ in front of these invariants. The first seven invariants used here respect the symmetry $ H \rightarrow - H $ and $\tilde H \rightarrow - \tilde H $, while the last two break this symmetry. These two  also break the symmetry under pure phase transformations. According to our vacuum structure both will contribute because of the presence of the $b$ terms in (\ref{b}). 
The total potential $ V = V_0 + V_S $ should now provide non-zero masses for all fields except the Goldstone ones.

As a first try one could set the small $b's$ to zero and needs then only the first seven invariants in (\ref{VS}).  A very appealing mass spectrum is achieved  with all other masses lying above the (almost) degenerate twins.    However, even the tiniest change of  $b$  in $\langle H \rangle $ away from zero destroys this picture. A fatal discontinuity is present changing abruptly the positivity of the mass matrix \footnote{I am very much indebted to my colleague Werner Wetzel who advised me about the multiple occurrence of these shrill discontinuities when proposing the vev of invariants.}.

Let us then use the vacuum structure (\ref{b}) where the continuous parameter $b$ appears in both matrix fields. In fact, this choice appears to be the only one for which the potential formed from the $9$ invariants allows for a positive definite mass matrix with masses above the (almost) degenerate twins.   As free parameters one can use $r_1$, $r_2$ and $r_7$. The requirement of vanishing first derivatives of $V$  - after shifting the fields - fixes then the remaining parameters   $r_3$, $r_4$, $r_5$ , $r_6$ , $r_8$ and $r_9$.
 For the case $1$ (the plus value in (\ref{b})) one obtains
  \begin{eqnarray}
 \label{r+}
 r_3 = r_1,~~r_4=r_2,~~r_9  = - r_8,  ~~ r_5= - 2 (r_1+r_2) - \frac{2 r_1 v_1^2}{b^2+M^2}~~\quad\quad ~~~~\\ 
 \nonumber
r_6= -\frac{2 (b^2 M^4 (b^2 + M^2) r_7 + 
   b^2 M^2 (b^2 + M^2) r_1 v_1^2 - (b^4 r_1 + b^2 M^2 (r_1 + r_7) + 
      M^4 (r_1 + r_7)) v_1^4 + r_1 v_1^8)}{
 M^2 (b^2 + M^2) (b^2 M^2 - v_1^4)} \\  
 \nonumber
r_8 = -\frac{2 r_1 v_1^3 (b^2 + M^2 - 2 v_1^2) (b^2 + M^2 + v_1^2)}{M (b^2 + M^2) (b^2 M^2 - v_1^4)}
   \end{eqnarray}
 
 The eigenvalues and eigenvectors of the corresponding mass matrix can be solved numerically. Using $ M=10^{13}~$GeV, ~$v_1= 123~$GeV,
 $b= 4~$GeV, $r_1=10^{-2}$, ~$r_2 = 4$,~$r_7 = 5$ one obtains the following mass values:
 \begin{eqnarray}
  \label{mass2} 
123.1, ~124.2, ~491~(2\times),~ 493~(2\times),~535~(4\times),~552~(2\times),\\ 
\nonumber 
756, ~ 778, ~  2 \cdot 10^{13} ~(2\times), ~ 4.5 \cdot 10^{13} ~(4\times), ~ 6.4 \cdot 10^{13} ~~\text{GeV}.~~\quad~~~
\end{eqnarray}
It is seen that  the two lowest states, the twins, are still almost degenerate even though the next higher states are considerably heavier. The twins are mixtures  of the fields $h^1_1$ and $\tilde h^1_1$ of about equal magnitude  i.e. the field which couples to fermions and the field not coupled to fermions appear with the same strengths in each state. There is very little admixture of other neutral fields. The next 2  states ($3$ and $4$) are strictly degenerate according to $CP$ invariance: a positively charged and a negatively charged boson. Both are mixtures with equal magnitude from fields of $H$ and $\tilde H $.  The part from $H$ belongs mainly to  the $SU(2)_L $ doublet  from the third column of $H$, the part from  $\tilde H$ belongs mainly to the doublet from the second column of this matrix field. The next $2$ states ($5$ and $6$) belong to the same doublets and  have therefore similar masses. The scalars $7$ to $10$ belong to the doublets of the first column again. The last two of these are charged and strictly degenerate. 

If we increase the parameter $r_1$ by a factor $2$  the mass of the heavier of the twins is increased making the mass difference of the twin states about $2$~GeV. A change of the numerical value of $b$ is  of little influence. Therefore, the model appears to be a good example for a model with  a twin structure of the lowest scalar states with properties which are not easily distinguishable from a pure standard model Higgs. One has a smooth behavior when changing the parameters within a large region.   However, if one changes  the vacuum structure by adding even a tiny new element, for instance $\langle \tilde h^3_3 \rangle $,  the mass matrix changes abruptly to an unphysical, not positive definite form. Because of these discontinuities it is necessary to say that the twin picture can only be maintained if there is a very strict relation between the vevs of the two multiplets like the one suggested in (\ref{b}). 

If the Higgs boson is not a twin, one can still have the superposition of the two types of fields and near degeneracies of states of higher mass. The sign of $b$ in the vacuum structure of $\tilde H $ could be negative, 
an option  we called case $2$. The relations between the $r$ values change and another range of input parameters $r_1$, $r_2$ and $r_7$ can be used. In spite of $v_2 = v_1$  the lowest state is then non-degenerate and the next state much higher in mass.  Still this boson is an almost equal superposition of the fields $h^1_1$ and  $ \tilde h^1_1$. Thus it is not a usual Higgs boson. It can soon be experimentally excluded because its  production cross-section is lower than the one of a standard Higgs boson.

 In a modified form of case 2 one may take $v_2$ different from $v_1$, close to zero. As an example we take $M=10^{13}$,~$b=4$~GeV, but $v_1 \simeq v~ $ and $v_2\simeq 5$~GeV to be slightly bigger than $b$. The corresponding value for $c_1$ is $c_1\simeq \frac{1}{2}.$ The lowest boson has then again a mass of $\approx 123~$GeV.  Choosing  $r_1 = r_2 = 4$ and $r_7=2\cdot 10^{-3} $ this state is not degenerate and the corresponding field is to $99\% $ the field $h^1_1.$  In this example the next $4$ states have almost the same mass of $984$~GeV (the two charged ones with identical masses) and are all equal weight superpositions with regard to the two fields of different couplings to fermions. In case 1 and in case 2 the bosons $3$ and $4$ have opposite electromagnetic  charges and are strictly degenerate by $ CP$ invariance.
 
The clarification whether the Higgs boson is a single resonance or a twin, or perhaps a degenerate state of a two-Higgs-doublet model  \cite {Ha}, requires more experimental data and a very detailed analysis by experts.

\section{Conclusions}\label{sec:4}

By extending the Glashow-Weinberg-Salam group to  $SU(3)_L\times SU(3)_R \times SU(3)_C $ as  in 
\cite{BZ, BS} the scalar sector consists of two $(3^*, ~3 ,~1 )$ multiplets $H$ and $\tilde H$ where only the fields in $H$ have a Yukawa coupling to quarks and leptons. The vacuum expectation values of $H$ and $\tilde H$ contain, besides the low mass elements responsible for the electro-weak symmetry breaking, also elements having a very high scale. A phenomenological potential has been constructed which reproduces these vevs. This invariant potential is taken to be of the Coleman-Weinberg type \cite{CW} and thus has no tree-level mass term. 
It is suggestive to assume a relation between the vev's of $H$ and $\tilde H$, in particular, to take the low scale elements $v_1$ in $H$ and $v_2$ in $\tilde H$  to be equal. Using at first only very simple $SU(3)_L\times SU(3)_R $ invariants, this assumption has the consequence that the two lowest scalar states are degenerate mass eigenstates with masses near  $\frac{v}{\sqrt{2}}= 123$~GeV. The two states are mixtures of a Higgs field with conventional properties and a Higgs field which can couple to fermions only via gauge vector bosons. Problematic for maintaining the degeneracy is the necessity  of including more invariants in order to get appropriate masses for the remaining fields. We considered the possibility for a one-to-one correspondence of the vevs of both multiplets and of their determinants. In the main case considered the twin structure can survive and the mass spectrum can be chosen to be in accord with experimental constraints.
We pointed out that by constructing potentials from vacuum expectation values discontinuities can appear which can only be tolerated if very strict relations between the vevs of the two multiplets hold. Nevertheless, it may be worthwhile to look for a twin structure of the Higgs boson found at the LHC.

Let me add a word on the quadratic divergencies. In the standard model - at least to one loop - the quadratic divergence can be viewed as being solely a problem for the vacuum 
expectation value of the Higgs field. The quadratic divergence of this quantity does not affect other particle properties and only indirectly the particle masses due to their couplings to the Higgs field.  Vacuum expectation values (like the cosmological constant) are not yet understood. But it is well known that by fine tuning the vacuum expectation value to its experimental value, or by subtracting the relevant tadpole graphs, or by assuming their cancellation at a very high scale \cite{old,Wet}, these divergencies have no further consequences for the particle spectrum. 
In the present phenomenological approach the only input are vev's - which themselves are certainly affected by quadratic divergencies - but are taken to be fixed.  Thus one can expect,  that in attempts of this type, the particle spectrum and other particle properties are not influenced by our non-understanding of vacuum expectation values and their quadratic divergencies.

\hspace{0.8cm}

{\bf Acknowledgment}\\
It is a pleasure to thank Ulrich Ellwanger, Tilman Plehn, David Lopez-Val and in particular Werner Wetzel for helpful discussions.
\vs{0.2cm}

\bibliographystyle{unsrt}

\begin{thebibliography}{99}



\bibitem{BSHiggs}
B. Stech, arXiv: hep-ph/1012.6028 (2010);\\
B. Stech Phys.Rev. D {\bf{86}}, 055003 (2012);
\bibitem{LHC}
G. Aad et al. [ATLAS Collaboration], Phys. Lett. B 716 (2012) 1;S
Chatrchyan et al. [CMS Collaboration], Phys. Lett. B 716 (2012) 30.

Most recent experimental updates from the winter conferences (IHEP):\\
The ATLAS collaboration, ATLAS-CONF-2012-158, ATLAS-CONF-2012-160, ATLAS-CONF-2012-161, ATLAS- CONF-2012-162, ATLAS-CONF-2012-163; \\The CMS collaboration, CMS-PAS-HIG-12-042, CMS-PAS-HIG-12-043, CMS-PAS-HIG-12-044, CMS-PAS-HIG-12-046, CMS-PAS-HIG-12-051,  CMS-PAS-HIG-12-041.\\
The ATLAS collaboration, ATLAS-CONF-2012-127; \\
The CMS Collaboration, CMS-PAS-HIG-12-045.

\bibitem{Schue}
T. Schuecker, arXiv: 0708.3344 v8 (2011);
\bibitem{Trini}
 Y.  Achiman and B. Stech, in Advanced Summer Institute on New Phenomena in Lepton and Hadron
Physics, eds. D. E. C. Fries and J. Wess (Plenum, New York, 1979);]\\
S. L. Glashow, in Fifth Workshop on Grand Unification, ed. K. Kang, H. Fried, and P. Frampton
(World Scientific, Singapore, 1984), p. 88;
\bibitem{TriniPak}
K.S. Babu, X.G. He and S. Pakvasa, Phys. Rev. D 33, 763 (1986);
\bibitem{E6}
F.~Gursey, P.~Ramond and P.~Sikivie,
  Phys.\ Lett.\  B {\bf 60} (1976) 177;\\
Y.~Achiman and B.~Stech,
 Phys.\ Lett.\  B {\bf 77} (1978) 389;\\
Q.~Shafi,
 Phys.\ Lett.\  B {\bf 79} (1978) 301;\\
R.~Barbieri, D.~V.~Nanopoulos and A.~Masiero,
  Phys.\ Lett.\  B {\bf 104} (1981) 194;
\bibitem{BZ}
  B. Stech and Z. Tavartkiladze,
  Phys. Rev. D {\bf 77} (2008) 076009,
  hep-ph/0311161];
\bibitem{BS}
B. Stech,
Fortsch.Phys. {\bf 58} : 692-698 (2010),
hep-ph/1003.0581;
\bibitem{CW}
S. Coleman and E. Weinberg Phys. Rev. D{\bf7} 1888 (1973);
\bibitem{Ha}
P.M. Ferreira, R. Santos, H.E. Haber and J.P. Silva,
arXiv: hep-ph/1211.3131 v4
\bibitem{old}
B. Stech,
arXiv: hep-ph/9811233;
hep-ph/9905357
\bibitem{Wet}
Ch. Wetterich
arXiv: hep-ph/1112.2910, v3;



\end{thebibliography}

 \end{document}